\documentclass{article}

\usepackage{arxiv}

\usepackage[utf8]{inputenc} 
\usepackage[T1]{fontenc}    
\usepackage{hyperref}       
\usepackage{url}            
\usepackage{booktabs}       
\usepackage{amsfonts}       
\usepackage{nicefrac}       
\usepackage{microtype}      
\usepackage{lipsum}
\usepackage{graphicx}
\graphicspath{ {./images/} }

\usepackage{amssymb}
\usepackage{amsmath}

\title{Comment on "An efficient code to solve the Kepler equation. Elliptic case"
}

\author{
	Daniele Tommasini$^{1}$\thanks{E-mail:daniele@uvigo.es (DT)}
	and David N. Olivieri$^{2,3}$
	\\
	$^{1}$Applied Physics Department, School of Aeronautic and Space Engineering, \\
	Universidade de Vigo, As~Lagoas s/n, 32004 Ourense, Spain\\
$^{2}$Computer Science Department, School of Informatics (ESEI), \\
Universidade de Vigo, As Lagoas s/n, 32004~Ourense,~Spain\\
$^{3}$Centro de Intelixencia Artificial, La Molinera, s/n, 32004 Ourense, Spain
}

\begin{document}
	\maketitle
	\begin{abstract}
In a recent MNRAS article, Raposo-Pulido and Pelaez (RPP) designed a scheme for obtaining very close seeds for solving the elliptic Kepler Equation with the classical and the modified Newton-Rapshon methods. This implied an important reduction in the number of iterations needed to reach a given accuracy. However, RPP also made strong claims about the errors of their method that are incorrect. In particular, they claim that their
accuracy can always reach the level $\sim5\varepsilon$, where $\varepsilon$ is the machine epsilon (e.g. $\varepsilon=2.2\times10^{-16} $ in double precision), and that this result is attained for all values of the eccentricity $e<1$ and the mean anomaly $M\in[0,\pi]$, including for $e$ and $M$ that are arbitrarily close to $1$ and $0$, respectively. However, we demonstrate both numerically and analytically that any implementation of the classical or modified Newton-Raphson methods for Kepler's equation, including those described by RPP, have a limiting accuracy of the order $\sim\varepsilon/\sqrt{2(1-e)}$. Therefore the errors of these implementations diverge in the limit $e\to1$, and differ dramatically from the incorrect results given by RPP. Despite these shortcomings,  the RPP method can provide a very efficient option for reaching such limiting accuracy. We also provide a limit that is valid for the accuracy of any algorithm for solving Kepler equation, including schemes like bisection that do not use derivatives. Moreover, similar results are also demonstrated for the hyperbolic Kepler Equation.  The methods described in this work can provide guidelines for designing more accurate solutions of the elliptic and hyperbolic Kepler equations. 
	\end{abstract}

	

\section{Introduction}

In the non-relativistic two body approximation, the time evolution of two celestial bodies moving in a bound orbit in each other's gravitational field  can be computed by solving the famous elliptic Kepler Equation (hereafter KE)
\begin{equation}
    M = E - e \sin E,
    \label{eq:Kepler}
\end{equation}
where $e$ is eccentricity, $E$ is the eccentric anomaly describing the instantaneous angular position,  
and $M$ is the mean anomaly, an angular measure of the time elapsed since a given passage from periapsis \cite{Roy2005}[Chap. 4]. 

For any fixed values of $e$ and $M$, solving the KE  for $E$ is equivalent to obtaining the root of the function
\begin{equation}
    f(E) \equiv E - e \sin E - M.
        \label{eq:def f}
\end{equation}
A common approach is to use the Classical or Modified Newton-Raphson  methods (CNR and MNR, respectively)
\cite{Danby1983,Conway1986,Gerlach1994,Fukushima1996,Palacios2002,Feinstein2006,Mortari2014,Raposo2017,Lopez2017}. Such algorithms provide a sequence of approximations $E_n$ that are expected to converge to the solution of the KE if a sufficiently good first guess $E_0$ (also called the seed) is provided. 

In particular, Raposo-Pulido and Pelaez (RPP) \cite{Raposo2017} described three procedures for obtaining  seeds for the CNR and MNR methods that are very close to the solution. The first two of these procedures use piecewise polynomials, defined over either 12 or  23 intervals, respectively, and are numerically satisfactory when the eccentricity $e$ is not very close to $1$. The third procedure is shown to be a better choice in the singular corner of KE, i.e. for $e$ close to $1$, including for the numerically difficult region $M\lesssim (1-e)^{3/2}$. The authors (RPP) argue that the use of such seeds  significantly reduces the number of iterations needed by the CNR and MNR methods to achieve a given level of accuracy. These analytical first guesses proposed by RPP are valuable achievements, and they can be used safely to improve the CNR and MNR methods, provided the error analysis and the iteration stopping condition given in Ref. \cite{Raposo2017} are modified following the guidelines that will be provided in Section \ref{sec:methods}.

However, RPP  made strong claims about the errors of their methods that we demonstrate here are incorrect.
In particular, they claim that by using  their proposed seeds,  the CNR and especially the MNR methods  can reach an accuracy at the level of $\sim 5\varepsilon$ for every value of $M\in[0,\pi]$ and $e\in[0,1)$, where $\varepsilon$ is the machine epsilon, e.g. $\varepsilon= 2.2\times10^{-16}$ for double precision. Moreover, RPP claim that this precision is attained even for $e$ arbitrarily close to $1$ and for every value of $M$, including those arbitrarily close to 0, and in particular for $M\lesssim (1-e)^{3/2}$, a regime to which they dedicate ample space. As a consequence, they also claim that their code converges throughout a wider region of the $(e,M)$ plane, as compared to alternative algorithms for solving KE, and that they can provide the value of the solution of KE with 15 and 34 decimals in double and quadruple precision, respectively, for every value of $e$ and $M$. 

Here, we demonstrate that these assertions are incorrect.
In Section \ref{sec:methods}, we show that the results described in Ref. \cite{Raposo2017} do not reproduce the correct values of the errors that can actually be  obtained with a careful numerical computation using their method, when a scan of the critical region is included. Moreover, RPP assume that when the limiting accuracy is reached the absolute error on the function $f(E)$ of Equation (\ref{eq:def f}) is equal to the upper bound of machine precision $\varepsilon$ for the entire domain of interest. Such an assumption, however natural it may seem, is erroneous. As we prove both numerically and theoretically in Section \ref{sec:methods}, when the effect of the machine precision in the context of floating point round-off error \cite{Goldberg1991,Kincaid2000,Higham2002} is properly taken into account, the absolute error on $f(E)$ is of the order of $\varepsilon E$, not $\varepsilon$.

We describe how to consistently and correctly calculate the relative and absolute errors due to floating point round-off error. Such considerations are also valid in the singular corner. We then demonstrate, both numerically and analytically, that any implementations of the CNR or MNR methods for the KE, including those described by RPP, have an unavoidable limiting accuracy of the order $\sim\varepsilon/\sqrt{2(1-e)}$. Therefore the errors of these implementations diverge in the limit $e\to1$, in dramatic contrast with the results presented in Ref. \cite{Raposo2017}. We also argue that this limiting accuracy is expected to affect also other methods that use derivatives, such as inverse series \cite{Stumpff1968,Colwell1993,Tommasini2021}, or splines 
 \cite{Tommasini2020a,Tommasini2020b}, or divisions by differences of values of $f$ for points that are close, as occurs in the Secant method or Inverse Quadratic Interpolation appearing in Brent's scheme \cite{Brent1973}. Moreover, CORDIC-like methods for solving the KE have also a similar limiting accuracy \cite{Zechmeister2018,Zechmeister2021}. 

For more general methods, such as bisection \cite{Brent1973}, that are not based upon divisions by derivatives or differences of $f$ values, we also derive a universal analytical expression for the limiting accuracy that can be achieved in the solution of KE within a given machine precision. This result only depends on the structure of the equation and not on the method used to solve it. As such, it affects all the algorithms that have been proposed, or could ever be proposed in the future, to solve KE, providing a universal limit on the error, which is in any case much lower than that obtained for the CNR and MNR schemes.  Finally, we also demonstrate similar limits for the accuracy of the solution of the hyperbolic Kepler Equation.



\section{Methods}
\label{sec:methods}

\subsection{Definition of the errors of the RPP method}
\label{sec:error and example}

For any given values of $e$ and $M$, 
KE, i.e., $f(E) \equiv E - e \sin E - M=0$, can be solved iteratively, in such a way that the $n+1$-th approximation,  $E_{n+1}$, of the solution is obtained by the recurrence, 
\begin{equation}
    E_{n+1} = E_n +\Delta_n,
    \label{eq:iterations}
\end{equation}
with $\Delta_n$ fashioned in different forms  depending on the method chosen. Using the definition of $f(E)$, the expressions for $\Delta_n$ is \begin{equation}
    \Delta_n = -\frac{f(E_n)}{f'(E_n)},
\end{equation}
in the CNR algorithm, and
\begin{equation}
    \Delta_n = \frac{-2f(E_n)}{f'(E_n)+\left[\operatorname{sgn}f'(E_n)\right] \sqrt{|f'^2(E_n)-2f(E_n)f''(E_n)|}},
\end{equation}
in the MNR algorithm  \cite{Raposo2017},
where $\operatorname{sgn}f'(E_n)$ is $1$ ($-1$) for positive (negative) $f'(E_n)$.

When the series $E_n$ converges to the correct solution of KE, the value $\Delta_n$ is commonly used to evaluate the error affecting the solution $E_n$ at order $n$. This implies the expression for the absolute error given in Equation (20) of Ref. \cite{Raposo2017},
\begin{equation}
    \mathcal{E}_\text{abs} = \frac{| f(E_n) |}{| 1 - e \cos E_n|}.
    \label{eq:error_RPP}
\end{equation}
As RPP explain \cite{Raposo2017}[page 1707], in their implementation of the CNR and MNR algorithms the
iteration ends when $|f(E_n)|=|E_n-e\sin E_n - M|$ becomes smaller than the machine error $\varepsilon$. This procedure implies an uncertainty  $\varepsilon$ on the value of $|f(E)|$, and a resulting dependence of the absolute error  on $E$ as $\mathcal{E}_\text{abs} \simeq \varepsilon /| 1 - e \cos E| $. If this expression were correct, the error for small values of $E$ would be as large as $\mathcal{E}_\text{abs} \simeq \varepsilon /| 1 - e| $, and it would diverge for $e\to1$. However, RPP claim that the maximum value $\mathcal{E}_\text{max}$ of $\mathcal{E}_\text{abs}$ can be controlled to the level $\simeq10^{-15}$ rad for every value of $M$ (and hence of $E$) even in the ``singular corner'' in the limit $e\to1$ \cite{Raposo2017}[pages 1707, 1708, 1709 and 1712]. They also dedicate special effort to design a very accurate seed for the most critical values of $M$, namely $M\lesssim (1-e)^{3/2}$ and even $M\lesssim 0.001 (1-e)^{3/2}$ \cite{Raposo2017}[pages 1705 and 1706].  However, precisely in this regime,  $E$ would be of the order $E\lesssim (1-e)^{1/2}$, so that, once again, their expression for the maximum error would be of the order $\mathcal{E}_\text{max} =\max\mathcal{E}_\text{abs} \simeq \varepsilon /(1 - e) $ and would diverge for $e\to1$.

In order to clarify the origin of this contradiction, and determine the correct numerical values of the errors, we  implemented the numerical routines necessary to solve the KE using the CNR and MNR methods together with the three procedures for the seed (first guess) described in Ref. \cite{Raposo2017}. In code, we included a switch statement that selects the most accurate of such different seeds,  corresponding to the smallest ratio $|f(E_0)|/| 1- e\cos E_0| $ for each values of $e$ and $M$.

\subsection{A numerical example}
\label{sec:example}

\begin{table}
	\label{tab:1}
	\centering
	\caption{
	Results for the solution of KE using the MNR and CNR methods in double precision for $1-e= 10^{-8}$ and $M=1.589565129427894\times10^{-12}$ rad. The seed $E_0$ was computed with the algorithm for the critical corner given in Ref. \cite{Raposo2017}. 
	The values of $E_n$, $f(E_n)$, and the error $|\Delta_n | = | E_{n+1}-E_n|$ are listed for every order of iteration $n\le 9$. The last column shows the difference $|E_n-E_\text{bisection}|$, where 
    $ E_\text{b}\equiv E_\text{bisection}=0.0001257862775234476$ rad is the precise value
	obtained with the bisection method, which solved KE in this point with an accuracy of $\mathcal{E}_\text{bisection}=2.7\times10^{-20}$ rad.
	}
	\begin{tabular}{c}
	     MNR  \\
	\end{tabular}
	\begin{tabular}{c|c|c|c|c} 
		\hline
 $n$   &  $E_n$  & $f(E_n)$  & $| \Delta_n |$  & $|E_n-E_\text{b.}|$ \\
    &  (rad)  &  (rad) &  (rad) & (rad) \\
		\hline
0  &  0.0001257862777707024  &     2.0$\times 10^{-20}$  &     1.1$\times 10^{-12}$  &     2.5$\times 10^{-13}$ \\
1  &  0.00012578627665402832  &    -3.4$\times 10^{-20}$  &     1.9$\times 10^{-12}$  &     8.7$\times 10^{-13}$ \\
2  &  0.00012578627856397613  &     2.0$\times 10^{-20}$  &     1.1$\times 10^{-12}$  &     1.0$\times 10^{-12}$ \\
3  &  0.00012578627744730205  &    -7.1$\times 10^{-21}$  &     4.0$\times 10^{-13}$  &     7.6$\times 10^{-14}$ \\
4  &  0.00012578627784393893  &     2.0$\times 10^{-20}$  &     1.1$\times 10^{-12}$  &     3.2$\times 10^{-13}$ \\
5  &  0.00012578627672726485  &    -3.4$\times 10^{-20}$  &     1.9$\times 10^{-12}$  &     8.0$\times 10^{-13}$ \\
6  &  0.00012578627863721266  &     2.0$\times 10^{-20}$  &     1.1$\times 10^{-12}$  &     1.1$\times 10^{-12}$ \\
7  &  0.00012578627752053858  &    -7.1$\times 10^{-21}$  &     4.0$\times 10^{-13}$  &     2.9$\times 10^{-15}$ \\
8  &  0.00012578627791717546  &     2.0$\times 10^{-20}$  &     1.1$\times 10^{-12}$  &     3.9$\times 10^{-13}$ \\
9  &  0.00012578627680050138  &    -3.4$\times 10^{-20}$  &     1.9$\times 10^{-12}$  &     7.2$\times 10^{-13}$ \\
		\hline
	\end{tabular}
	\begin{tabular}{c}
	     CNR  \\
	\end{tabular}
	\begin{tabular}{c|c|c|c|c} 
		\hline
 $n$   &  $E_n$  & $f(E_n)$  & $| \Delta_n |$  & $|E_n-E_\text{b.}|$  \\
    &  (rad)  &  (rad) &  (rad) & (rad) \\
		\hline
0  &  0.0001257862777707024  &     2.0$\times 10^{-20}$  &     1.5$\times 10^{-12}$  &     2.5$\times 10^{-13}$ \\
1  &  0.00012578627625739144  &    -3.4$\times 10^{-20}$  &    1.9$\times 10^{-12}$  &     1.3$\times 10^{-12}$ \\
2  &  0.0001257862777707024  &     2.0$\times 10^{-20}$  &     1.1$\times 10^{-12}$  &     2.5$\times 10^{-13}$ \\
3  &  0.00012578627625739144  &    -3.4$\times 10^{-20}$  &     1.9$\times 10^{-12}$  &     1.3$\times 10^{-12}$ \\
4  &  0.0001257862777707024  &     2.0$\times 10^{-20}$  &     1.1$\times 10^{-12}$  &     2.5$\times 10^{-13}$ \\
5  &  0.00012578627625739144  &    -3.4$\times 10^{-20}$  &     1.9$\times 10^{-12}$  &     1.3$\times 10^{-12}$ \\
6  &  0.0001257862777707024  &     2.0$\times 10^{-20}$  &     1.1$\times 10^{-12}$  &     2.5$\times 10^{-13}$ \\
7  &  0.00012578627625739144  &    -3.4$\times 10^{-20}$  &     1.9$\times 10^{-12}$  &     1.3$\times 10^{-12}$ \\
8  &  0.0001257862777707024  &     2.0$\times 10^{-20}$  &     1.1$\times 10^{-12}$  &     2.5$\times 10^{-13}$ \\
9  &  0.00012578627625739144  &    -3.4$\times 10^{-20}$  &     1.9$\times 10^{-12}$  &     1.3$\times 10^{-12}$ \\		\hline
		\hline
	\end{tabular}
\end{table}

Table 1 shows the results of applying the MNR and CNR methods with the best RPP seed using double precision in an illustrative example, corresponding to $1-e=10^{-8}$ and $M= 1.589565129427894\times10^{-12}$ rad. The corresponding best seed, as expected from the discussion in Ref. \cite{Raposo2017}, is that obtained from their algorithm in the critical corner. In this case, this gives $E_0= 0.0001257862777707024$ rad. It can be seen that for both the MNR and CNR methods the values of the function $f(E_n)$ remain consistently below $|f(E_n)|\le 3.4\times 10^{-20}\ll \varepsilon$ iteration after iteration, including when $n=0$. In this case, according to Ref. \cite{Raposo2017}, no iteration would have been performed, and the output of the method would be $E_0$. However the error, as computed either with Equation (\ref{eq:error_RPP}) or with $\Delta_n$, 
cannot be reduced below $\mathcal{E}_\text{abs}\ge 1.1\times 10^{-12}$, without any improvement coming from additional iterations. The values of the solutions $E_n$ vary 
within a range $\mathcal{O}( 10^{-12})$ rad, 
thus contradicting the claim made in Ref. \cite{Raposo2017} that 
the error could  always be kept below the $10^{-15}$ rad level in double precision thanks to their more precise seed. In this example, the limit of the error that can be achieved with the CNR and MNR methods, even when they are enhanced with RPP seeds, is three orders of magnitude higher that the value given in Ref. \cite{Raposo2017}. Fortunately, it is also much smaller, by a factor $f(E_n)/\varepsilon \simeq 10^{-4}$,
than the error that could be expected for this value of $E$ by substituting the accuracy  $\varepsilon$ that they assume for $f$ in their Equation (20)---our Equation (\ref{eq:error_RPP}). Thus, this example shows that the seed designed in Ref. \cite{Raposo2017} is an excellent choice for the singular corner, since it is sufficient to achieve this limiting error bound $10^{-12}$ rad. However, the error in this region was not correctly reported by RPP.

In order to verify whether this limit on the accuracy is specific of the CNR and MNR methods, we computed the solution of KE for the values of $e$ and $M$ considered in this example using the bisection root search method \cite{Brent1973} on $f(E)$, stopping the iterations when no further improvement in the accuracy was obtained. 
The result, obtained after 66 bisection iterations between the initial endpoints $0$ and $\pi$, was the value $E_\text{bisection}=0.0001257862775234476$ rad with an error $\mathcal{E}_\text{bisection}=2.7\times10^{-20}$ rad. Since bisection only uses the values of the function $f(E)$, this can be thought to be the best accuracy that can be achieved with any method for these values of $e$ and $M$. This limiting accuracy, $\mathcal{E}_\text{bisection}$, is eight orders of magnitude more precise than that obtained with the CNR and MNR methods in this example. This ratio is due to the factor $1-e\cos E$  in Equation (\ref{eq:error_RPP}), which gives $\simeq 1.8\times10^{-8}$ for the value of $E$ considered in this example.

This precise result,  ${E}_\text{bisection}$, can also be used to obtain an independent measure of the error of the RRP method for the values of $e$ and $M$ considered in this example by computing the differences $|E_n-E_\text{bisection}|$, given in the last column of Table 1. It can be seen that increasing the order of the iterations does not produce convergence to this more precise solution, and the error of the RPP method, estimated with this comparison, is of the order of $10^{-12}$ rad, in agreement with the self-consistent error $\Delta_n$ computed only with the MNR or CNR methods.

Notice also that the relative error  of the solution obtained using the bisection method turns out to be $\mathcal{E}_\text{bisection}/E_\text{bisection} \simeq 2.1\times10^{-16}\simeq\varepsilon$. As discussed in Subsection \ref{sec:theoretical limit elliptical}, this is precisely the limiting relative error with which the variable $E$ can be defined using double precision floating point. Moreover, the value of $f(E)$ in this best approximation for the solution is $f({E}_\text{bisection})\simeq 2.0\times10^{-20}$ rad, which is roughly equal to the value of the error $\mathcal{E}_\text{bisection} $. As we shall discuss in Subsection \ref{sec:theoretical limit elliptical}, this fact is a consequence of the $E$ dependence of $f$.

\subsection{Theoretical estimate for the limiting accuracy of the CNR and MNR methods}
\label{sec:theoretical limit elliptical}

A key question is to quantify the effect $\delta f$ of the machine error on $f$.
Since the central value of $f(E)$ for a solution of KE is zero, it could be thought that such error would be $\delta f=\varepsilon$. This was the assumption made in Ref. \cite{Raposo2017}. 
In this case, 
as we have discussed, Equation (\ref{eq:error_RPP}) would imply that the maximum error would diverge as $\varepsilon/(1-e)$. However, the effect of the machine precision on $f(E)$ is a consequence of its 
dependence on $E$. Let $\delta E$ be the absolute uncertainty on the variable $E$ due to the machine precision, to be distinguished from the error $\mathcal{E}$ on the solution of KE. As for any variable, the \emph{relative} uncertainty on $E$ is
$\frac{\delta E}{E}\simeq \varepsilon$, as discussed in classic texts on numerical computations and machine precision \cite{Kincaid2000}[pp.44-46] and \cite{Higham2002}[pp.37-38]. 
Therefore
\begin{equation}
    \delta E \simeq \varepsilon E,
\end{equation}
and this will also be a universal lower limit on the accuracy $\mathcal{E}_\text{abs}\gtrsim \varepsilon E$  for solving KE, which can be reached using bisection as discussed in the example of subsection \ref{sec:example}. However, the limit for the error level of the CNR and MNR methods, even with the best seed, will be higher.
For such methods, the effect of the round-off  error $\delta E \simeq \varepsilon E$ on $E$ will be propagated in the function $f(E)$, so that  $\delta f \simeq \delta E\simeq \varepsilon E$. In fact, we have seen that this is the case in the example considered in Subsection \ref{sec:example}.
Since when $f(E)\simeq0$ the value of $f$ oscillates within the error $\delta f$, we can substitute $f$ with this expression for $\delta f$ in Equation (\ref{eq:error_RPP}) and obtain,
\begin{equation}
   \mathcal{E}_\text{abs} \simeq \frac{\varepsilon\, E}{1-e\cos E}.
   \label{eq:limit-abs-error}
\end{equation}
This expression describes very well the numerical errors of the CNR and MNR methods. This agreement has already been seen in the example of Subsection \ref{sec:example}, and it will also be confirmed by the numerical analysis of the $M$ dependence of the errors (see Subsection \ref{sec:M-dependence} below).

The value $E_\text{max error}$ corresponding to the maximum error for a given value of the eccentricity can be obtained by setting to zero the derivative of Equation (\ref{eq:limit-abs-error}), so that $1-e(\cos E+ e E\sin E) =0 $. 
Since in the critical region $e$ is close to $1$, the maximum of $\mathcal{E}_\text{abs}$ is obtained for small values of $E$. By Taylor expanding the sine and cosine functions, we obtain the following approximation for $E_\text{max error} $,
\begin{equation}
    E_\text{max error} \simeq \sqrt{2(1-e)}.
    \label{eq:E max error}
\end{equation}
This implies a limiting accuracy
\begin{equation}
    \mathcal{E}_\text{lim}\simeq 
    \frac{\varepsilon}{\sqrt{2(1-e)}}.
    \label{eq:limit-error-e}
\end{equation}

We can now estimate the range of values of $e$ and $M$ for which the results for the accuracy $\mathcal{E}_\text{RPP}$  given in Ref. \cite{Raposo2017} can be correct.
The condition $\mathcal{E}_\text{lim}\le \mathcal{E}_\text{RPP}$ implies,
\begin{equation}
    e\lesssim    1 - \frac{1}{2}\left(\frac{\varepsilon}{\mathcal{E}_\text{RPP}}\right)^2.
    \label{eq:switch-e}
\end{equation}
In double precision, $\varepsilon =2.2\times10^{-16}$ and  $\mathcal{E}_\text{RPP} =10^{-15}$, this gives $e \lesssim 0.976$. For values of $e$ higher than these limits, the errors given in Ref. \cite{Raposo2017} are incorrect, except in a $M$ region that can be obtained by solving the inequality $\mathcal{E}_\text{abs}\le \mathcal{E}_\text{RPP}$, with $\mathcal{E}_\text{abs}$ given by Equation (\ref{eq:limit-abs-error}). Since the value of $E $ in which $\mathcal{E}_\text{abs}= \mathcal{E}_\text{RPP}$ is not small, this equation has to be solved numerically. For $e\to 1$, the solution is  $E\gtrsim 0.44$ and $M\gtrsim 0.014$ rad. In other words, the errors given in Ref. \cite{Raposo2017} are expected to be unreliable when both $e \gtrsim 0.976$ and  $M\lesssim 0.014$ rad, which is the core of the singular region.

These theoretical results apply to both the CNR and MNR methods, and they are confirmed by the numerical example discussed in Subsection \ref{sec:example} and by the analysis of the $M$ dependence of the errors that will be presented in Subsection \ref{sec:M-dependence}.

\subsection{$M$ dependence of the errors}
\label{sec:M-dependence}

\begin{figure}
	\includegraphics[width=\columnwidth]{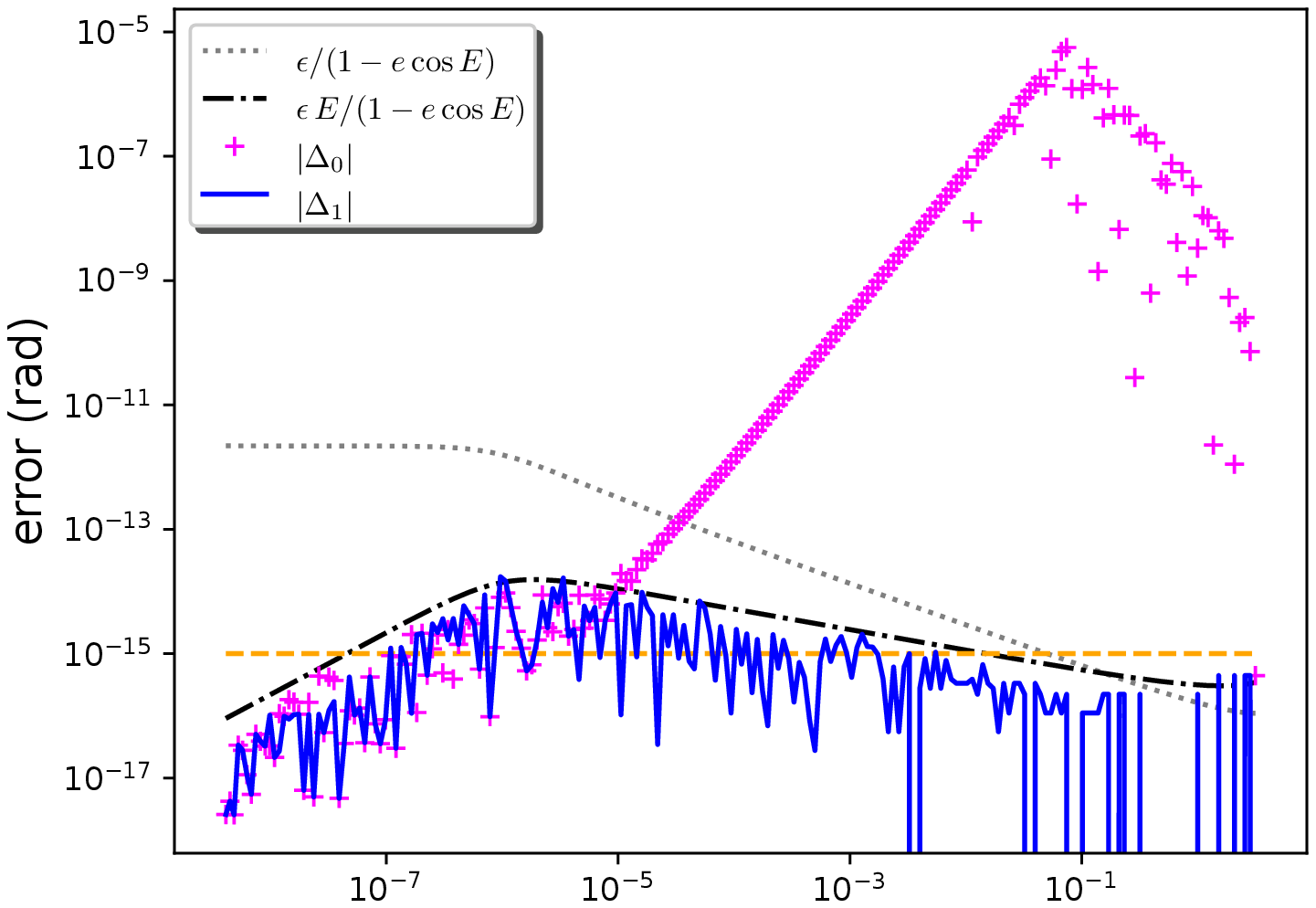}
	\caption{$M$ dependence of the errors for solving KE in double precision for $1-e= 10^{-4}$. 
	The  errors $|\Delta_0| = | E_1 - E_0| $ (magenta crosses) and $|\Delta_1| = | E_2 - E_1| $ (continuous blue line) correspond to the zeroth and first order solutions of KE, respectively, obtained with the MNR method of Ref.  \cite{Raposo2017} using their best seed.	
	The horizontal dashed orange  line represents the underestimated maximum error given there. 
	The dot dashed black curve was obtained using our theoretical expression for the limiting accuracy, Equation
	(\ref{eq:limit-abs-error}). No further reduction of the error is obtained by increasing the number of iterations.
	}
	\label{fig:figure1e-4}
\end{figure}

\begin{figure}
	\includegraphics[width=\columnwidth]{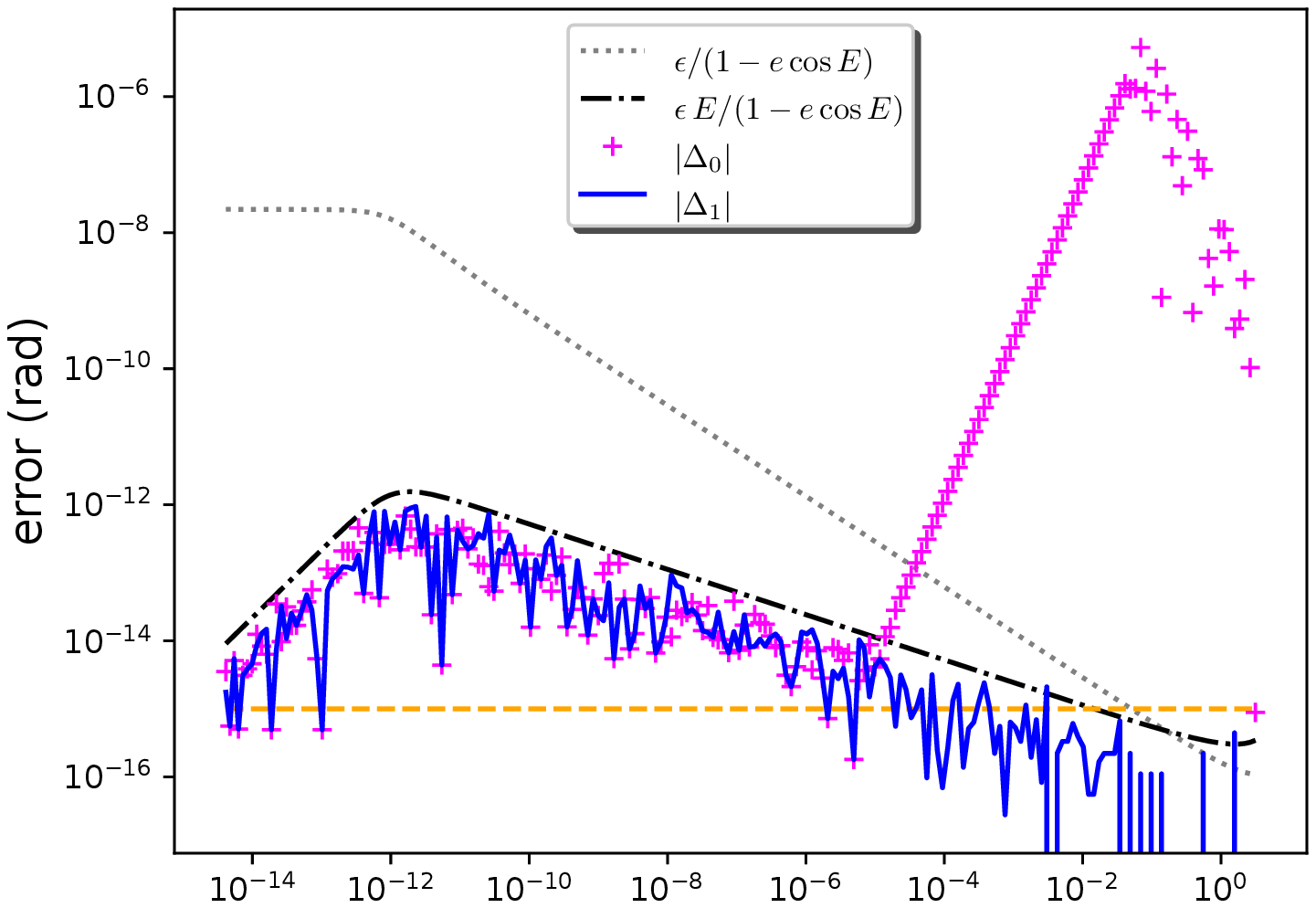}
	\caption{$M$ dependence of the errors  for solving KE in double precision for $1-e= 10^{-8}$. The  errors $|\Delta_0| = | E_1 - E_0| $ (magenta crosses) and $|\Delta_1| = | E_2 - E_1| $ (continuous blue line) correspond to the zeroth and first order solutions of KE, respectively, obtained with the MNR method of Ref.  \cite{Raposo2017}  using their best seed. The horizontal dashed orange  line represents the underestimated maximum error given there. The dot dashed black curve was obtained using our theoretical expression for the limiting accuracy, Equation
	(\ref{eq:limit-abs-error}). No further reduction of the error is obtained by increasing the number of iterations.
	}
	\label{fig:figure1e-8}
\end{figure}

\begin{figure}
	\includegraphics[width=\columnwidth]{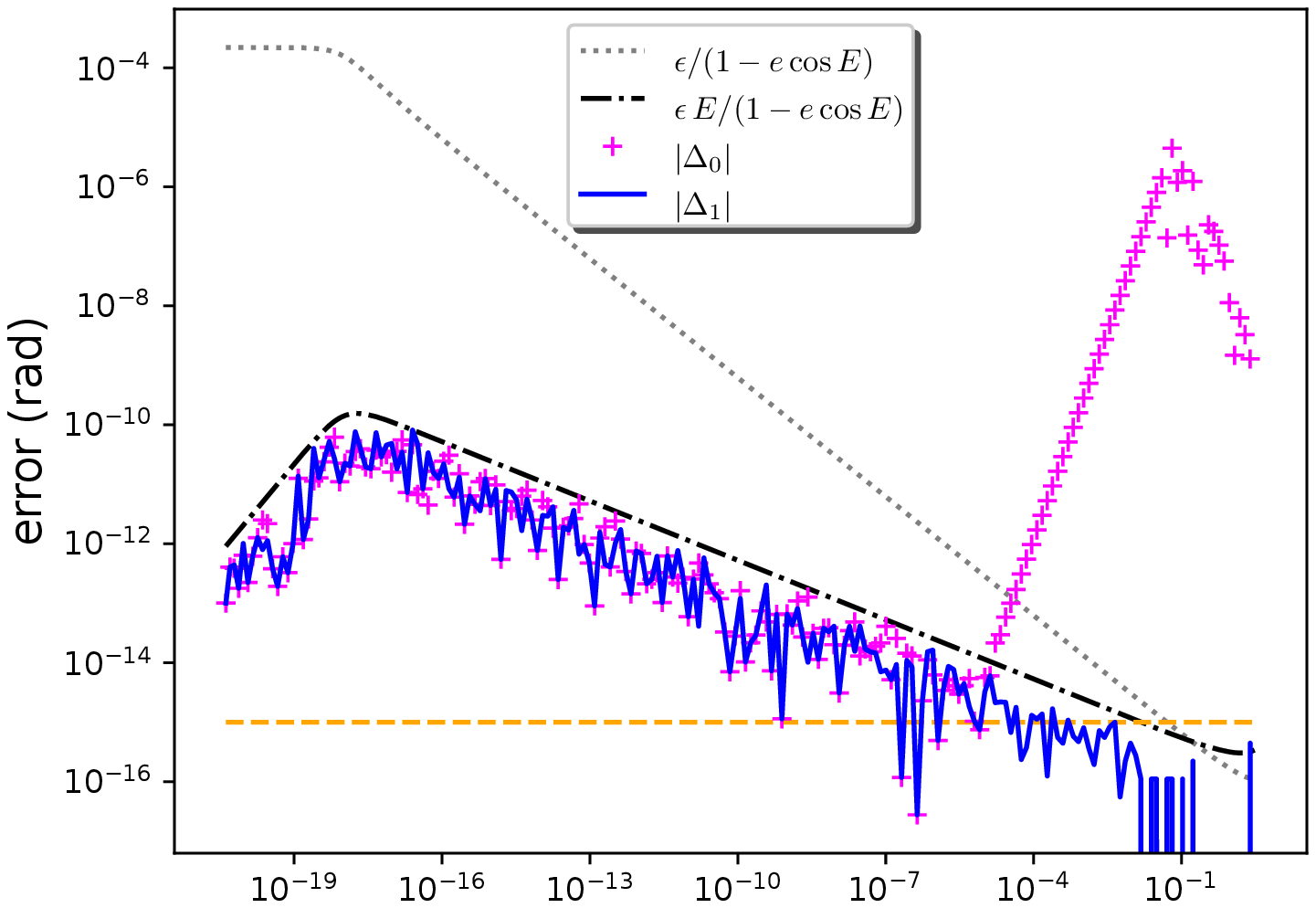}
	\caption{$M$ dependence of the errors for solving KE in double precision for $1-e= 10^{-12}$. 
The  errors $|\Delta_0| = | E_1 - E_0| $ (magenta crosses) and $|\Delta_1| = | E_2 - E_1| $ (continuous blue line) correspond to the zeroth and first order solutions of KE, respectively, obtained with the MNR method of Ref.  \cite{Raposo2017}  using their best seed.	
	The horizontal dashed orange  line represents the underestimated maximum error given there. 
	The dot dashed black curve was obtained using our theoretical expression for the limiting accuracy, Equation
	(\ref{eq:limit-abs-error}). No further reduction of the error is obtained by increasing the number of iterations.
	}
	\label{fig:figure1e-12}
\end{figure}

Figures \ref{fig:figure1e-4}, \ref{fig:figure1e-8} and \ref{fig:figure1e-12}  show the $M$ dependence of the error distribution for solving KE with the MNR algorithm starting from the RPP seed and using double precision. These error distributions were computed numerically from the value of $\Delta_n$ for three different values of $e$ in the singular corner. It can be seen that in all cases the numerical results agree with the theoretical predictions from Equation (\ref{eq:error_RPP}). The errors oscillate around zero, so that the global accuracy is the maximum error at the absolute peak.

We verified that only one iteration is sufficient to attain the maximum allowed accuracy for the MNR method in double precision, in agreement with Ref. \cite{Raposo2017}, since further iterations do not significantly reduce the errors. One or two additional iterations may be needed with the CNR algorithm. This is a remarkable achievement of the seeds designed in Ref. \cite{Raposo2017}, which are very accurate especially in the singular region, for $e\to1$ and $M\lesssim (1-e)^{3/2}$. 
However, it can be seen that the error is larger than the value $10^{-15}$ rad given in Ref. \cite{Raposo2017}, and the discrepancy grows the closer is $e$ to $1$. Fortunately, the correct error, while being larger than that given in Ref. \cite{Raposo2017} (indicated by the horizontal dashed orange line), is also significantly smaller than the dotted gray curve in the figures, representing
what could be expected from Equation (\ref{eq:error_RPP}) and the assumption of Ref. \cite{Raposo2017} that the error on $|f(E)|$ is $\varepsilon$.
Our theoretical prediction of Equation (\ref{eq:limit-abs-error}) for the limiting accuracy is represented by the dot-dashed curve in Figures \ref{fig:figure1e-4}, \ref{fig:figure1e-8},  \ref{fig:figure1e-12}. The agreement with the actual numerical errors is excellent.

\begin{figure}
	\includegraphics[width=\columnwidth]{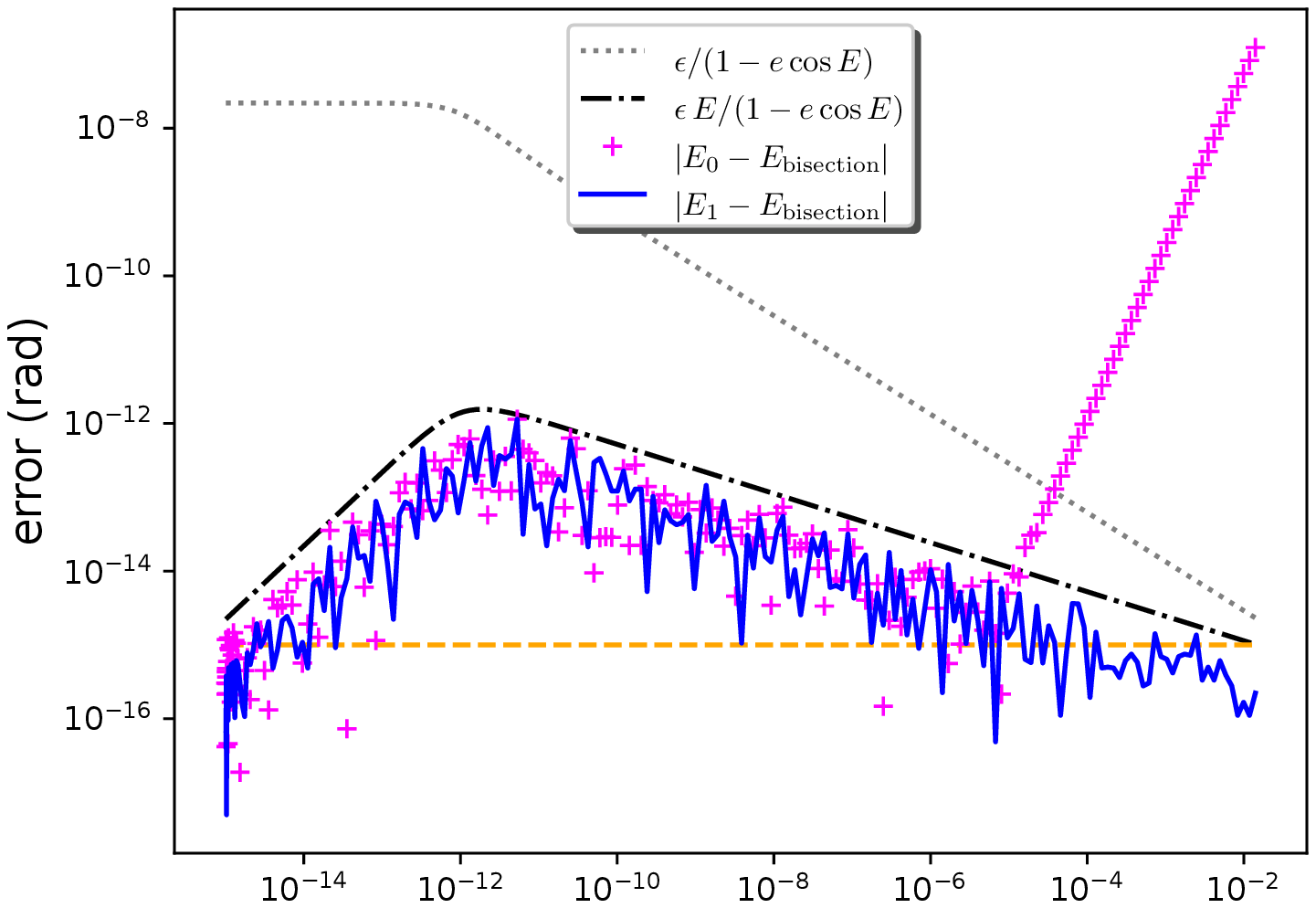}
	\caption{Alternative computation of the $M$ dependence of the errors  for solving KE in double precision for $1-e= 10^{-8}$. The  errors $|E_0-E_\text{bisection}|$  (magenta crosses) and $|E_1-E_\text{bisection}|$ (continuous blue line) correspond to the zeroth and first order solutions of KE, respectively, obtained with the MNR method of Ref.  \cite{Raposo2017} using their best seed.	$E_\text{bisection}$ is the precise solution obtained with the bisection method with tolerance $\mathcal{E}_\text{bisection} = 1.1\times 10^{-16}$ rad. No further reduction of the error is obtained by increasing the number of iterations. Only the errors in the critical region $M<0.015$ rad are shown. These results agree with those obtained with the direct computation of the errors of the method shown in Figure \ref{fig:figure1e-8}.
	}
	\label{fig:figure4e-8}
\end{figure}

An independent computation of the errors of the RPP method can be obtained by comparing its results at a given order of iteration, $E_n$, with the very precise values $E_\text{bisection}$ obtained with the bisection method. Figure \ref{fig:figure4e-8} shows the differences  $|E_0-E_\text{bisection}|$ and $|E_1-E_\text{bisection}|$ corresponding to the zeroth and first order solutions of KE, respectively, obtained with the MNR method of Ref.  \cite{Raposo2017} for $1-e= 10^{-8}$. Only the results in the critical region $M<0.014$ rad are shown, corresponding to $E\lesssim 0.44$ as seen in Subsection \ref{sec:theoretical limit elliptical}. 

From the discussion in Subsection \ref{sec:theoretical limit elliptical}, we know that in such an interval the bisection method can attain convergence with an input tolerance as low as $\sim0.44\varepsilon$. In practice, the values of  $E_\text{bisection}$ used in the figure were
obtained with input tolerance $\mathcal{E}_\text{bisection} = 0.5\varepsilon=1.1\times 10^{-16}$ rad, so that their error can be neglected in comparison with that of the RPP method.  In any case, the resulting values of $|E_0-E_\text{bisection}|$ and $|E_1-E_\text{bisection}|$ agree with the errors obtained with the direct computation shown in Figure \ref{fig:figure1e-8}.
Again, since no further reduction of this difference is obtained by increasing the number of iterations, Figure \ref{fig:figure4e-8} also confirms that the limiting accuracy of the RPP method is very well described by our theoretical expression of Equation (\ref{eq:limit-abs-error}) (corresponding to the dot dashed black curve in the figure).

\begin{figure}
	\includegraphics[width=\columnwidth]{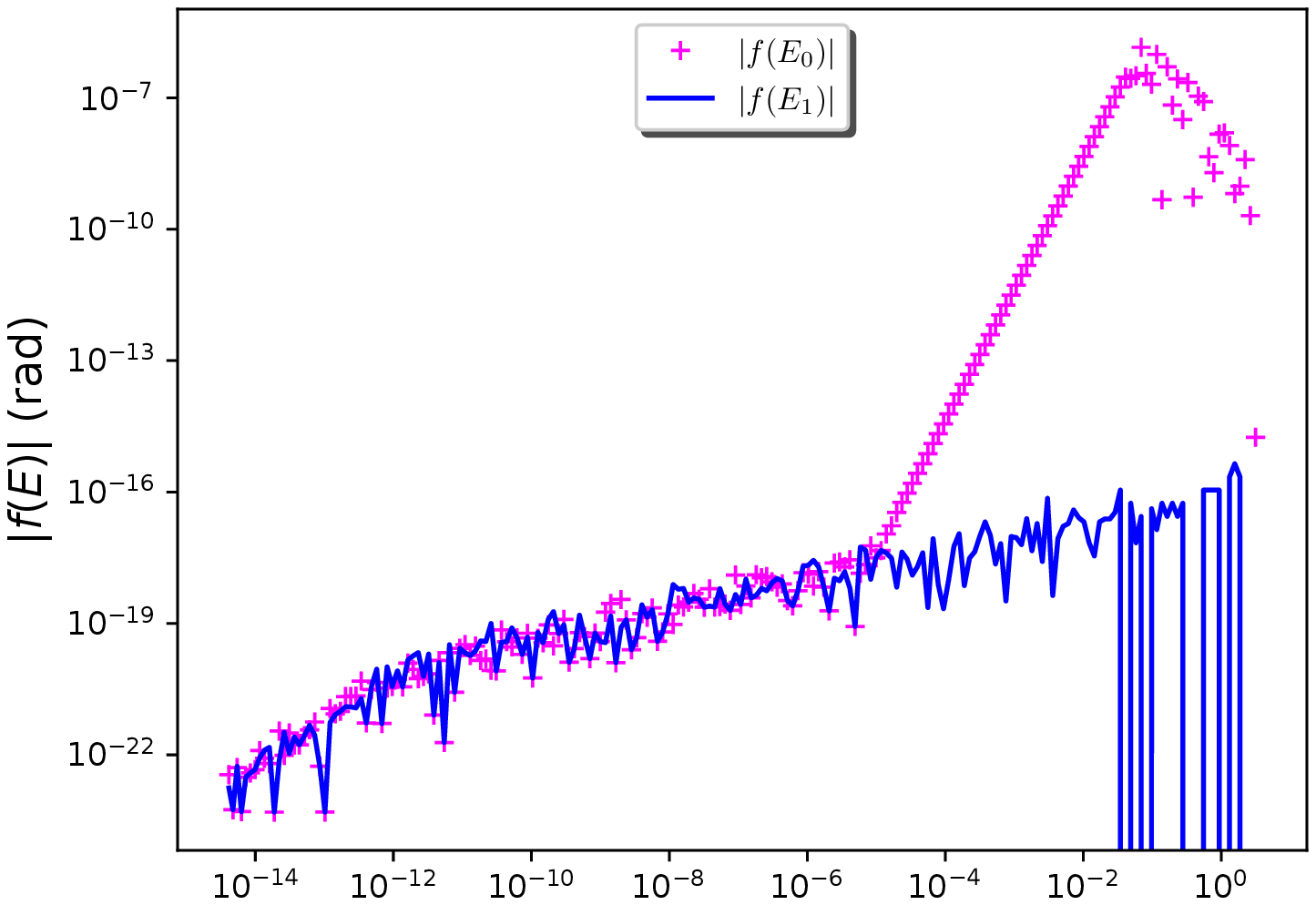}
	\caption{Values of $|f(E_0)|$ (magenta crosses) and $|f(E_1)|$ (continuous blue line) for the zeroth and first order solutions of KE, $E_0$ and $E_1$, obtained in double precision for $1-e= 10^{-8}$ with the MNR algorithm using the best seed given in Ref. \cite{Raposo2017}.
	}
	\label{fig:figure2e-8}
\end{figure}

The fact that the error can be kept much smaller than the level $\varepsilon/(1-e\cos E)$
for small $E$ requires that the value of $|f(E)| $, and the error affecting such value, should be controlled to be well below the machine epsilon $\varepsilon$ for small $E$, as we have also seen in the example of Subsection \ref{sec:example} and in the theoretical discussion of Subsection \ref{sec:theoretical limit elliptical}. This is indeed the case in general, as shown in Figure \ref{fig:figure2e-8} for $1-e=10^{-8}$ (similar results are obtained for different values of $e$ in the critical corner). In other words, for small $E$ the limiting accuracy can only be achieved when $|f(E)|$ is much smaller than $\varepsilon$. Therefore, in general, using the MNR of CNR methods and stopping the iteration when $|f(E) | $ is smaller than $\varepsilon$ does not guarantee that the limiting allowed accuracy is reached (besides the fact that such limit is higher than $10^{-15}$ rad). 

\begin{figure}
	\includegraphics[width=\columnwidth]{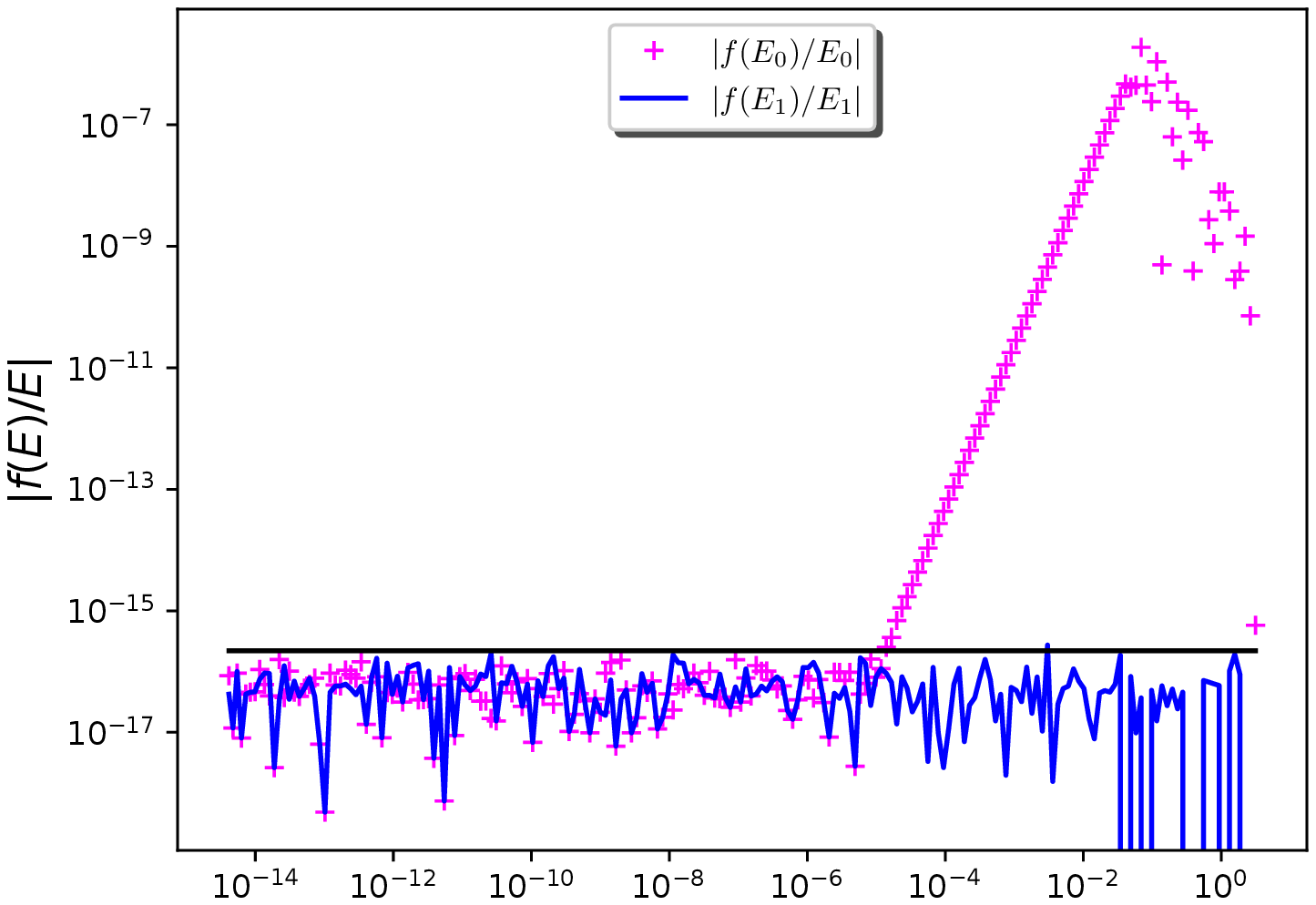}
	\caption{Values of $|f(E_0)/E_0|$ (magenta crosses) and $|f(E_1)/E_1|$ (continuous blue line) for the zeroth and first order solutions of KE, $E_0$ and $E_1$, obtained in double precision for $1-e= 10^{-8}$ with the MNR algorithm using the best seeds given in Ref. \cite{Raposo2017}. It can be seen that the values of $|f(E_1)/E_1|$ oscillate below the level $ \varepsilon$ (horizontal line) for every $M\in [0,\pi]$.
}
	\label{fig:figure3e-8}
\end{figure}

A better insight of the machine precision dependence  of $f(E)$ at the limiting accuracy can be obtained by plotting the ratio $|f(E)|/E$, as in Figure  \ref{fig:figure3e-8} (also corresponding to $1-e= 10^{-8}$). It can be seen that $|f(E)|/E\lesssim \varepsilon$, 
reaching the level $|f(E)|/E\simeq\varepsilon$ in the peaks of the oscillations. Therefore, it is the ratio $|f(E)|/E$ that becomes smaller than the machine precision at the limiting accuracy, rather than $|f(E)|$. In other words, the effect of the machine precision on the solution of KE induces an uncertainty $ \varepsilon E$ on $f(E)$. Again, this result agrees with the theoretical prediction given in Subsection \ref{sec:theoretical limit elliptical}.

\subsection{Theoretical limiting accuracy for the hyperbolic KE}
\label{sec:bisection}

Similar results can be obtained with the hyperbolic KE, describing the time dependence of the solution of the two body problem for eccentricity $e>1$. This equation can be written as $f(H)=0$ by defining the function
\begin{equation}
    f(H) = e\sinh H - H - M,
\end{equation}
where $M$ and $H$ are the hyperbolic mean and eccentric anomaly, respectively \cite{Roy2005, Gooding1988, Fukushima1997, Avendano2015, Raposo2018, Tommasini2021}. Both $M$ and $H$ vary in the interval $(-\infty,\infty)$. Since the error of NR methods and its generalizations are proportional to $|f(H)/f'(H)|$, they become singular for $e\to1$ and $|M|\ll1$, just like in the case of the elliptic KE. 

Again, the round-off  error $\delta H = \varepsilon H$ is expected to set a universal lower limit on the accuracy $\mathcal{E}_\text{abs}\gtrsim \varepsilon H$  for solving the hyperbolic KE, which can be reached using bisection.  However, the limit for the error level of the CNR and MNR methods, even with the best seed, will be higher.
For such methods, as in the case of the elliptic KE, the effect of the round-off  error $\delta H = \varepsilon H$ on $H$ will be propagated in the function $f(H)$, so that  $\delta f \simeq \delta H\simeq \varepsilon H$. Again, when $f(H)\simeq0$ the value of $f$ oscillates within the error $\delta f$, thus we can substitute $f$ with this expression for $\delta f$ in the expression for the error $\Delta_n$ of Newton-Raphson method and obtain,
\begin{equation}
   \mathcal{E}_\text{abs} \simeq \left|\frac{\varepsilon\, H}{e\cosh H - 1}\right|.
   \label{eq:limit-abs-error-H}
\end{equation}

The value $H_\text{max error} $ corresponding to the maximum error for a given value of the eccentricity can be obtained by setting to zero the derivative of Equation (\ref{eq:limit-abs-error-H}), so that $-1+e(\cosh H- H\sin H) =0 $. 
Since in the core of the critical region $e$ is close to $1$ and $H\ll1$, a Taylor expansion of the hyperbolic functions can be used to obtain the following approximation,
\begin{equation}
    H_\text{max error} \simeq \sqrt{2(e-1)}.
    \label{eq:E max error H}
\end{equation}
This implies a limiting accuracy
\begin{equation}
    \mathcal{E}_\text{lim}\simeq 
    \frac{\varepsilon}{\sqrt{2(e-1)}}.
    \label{eq:limit-error-e-H}
\end{equation}

We can now estimate the range of values of $e$ and $M$ for which the accuracy can be set at a certain level $\mathcal{E}$, for instance $10^{-15}$. The condition $\mathcal{E}_\text{lim}\le \mathcal{E}$ implies,
\begin{equation}
    e\lesssim    1 - \frac{1}{2}\left(\frac{\varepsilon}{\mathcal{E}}\right)^2.
    \label{eq:switch-e-hyper}
\end{equation}
In double precision, $\varepsilon =2.2\times10^{-16}$, so that  $\mathcal{E} =10^{-15}$ implies $e \gtrsim 1.024$. For values of $e$ lower than this limit, the errors will be higher, except in a $M$ region that can be obtained by solving the inequality $\mathcal{E}_\text{abs}\le \mathcal{E}$, with $\mathcal{E}_\text{abs}$ given by Equation (\ref{eq:limit-abs-error-H}). The numerical solution of this equation is  $H\gtrsim 0.53$ and $M\gtrsim 0.012$. Notice that these numbers are similar to those obtained for the elliptic KE since the Taylor expansions for small $E$ and $H$ for the two cases are similar, with the change $1-e\leftrightarrow e-1$.  

\begin{figure}
	\includegraphics[width=\columnwidth]{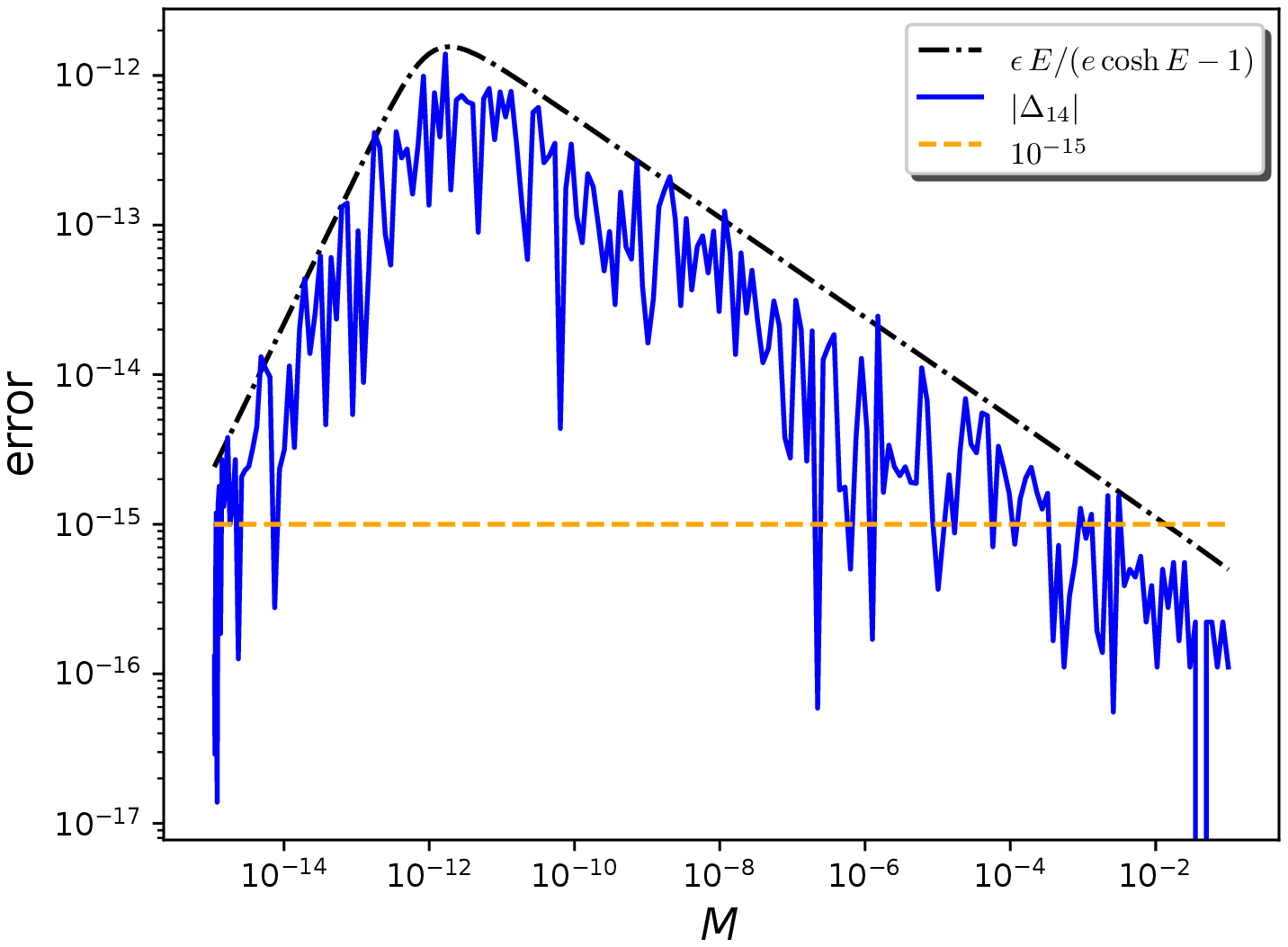}
	\caption{$M$ dependence of the errors for solving the hyperbolic KE in double precision for $e-1= 10^{-8}$. The continuous blue line corresponds to the values of the error $|\Delta_{14}| = | E_{15} - E_{14}| $ 	for the MNR algorithm with 14 iterations using the simple seed $H_0=M$, the horizontal dashed orange line representing the underestimated error given in Ref. \cite{Raposo2018}. The dot dashed black curve was obtained using our theoretical expression for the limiting accuracy, Equation
	(\ref{eq:limit-abs-error-H}). No further reduction of the error is obtained by increasing the number of iterations.
	}
	\label{fig:figure-he-8}
\end{figure}

Figure \ref{fig:figure-he-8} shows the errors obtained with the MNR algorithm for $e=1+10^{-8} $ in the region $M<0.1$ using double precision. In this case, $n=14$ iterations starting from the simple seed $H_0=M$ were used to achieve this result. Although other seed choices, such as that given in Ref. \cite{Raposo2018}, can reach such precision with a much smaller number of iterations, the fact that no improvement of the precision is found by increasing $n$ implies that the line of the local maxima of the error describe the numerical limiting accuracy. It can be seen that such limit is in excellent agreement with the theoretical expression for the minimum error level given by Equation (\ref{eq:limit-abs-error-H}), represented by the dot-dashed curve in the figure.

\subsection{Limit accuracy for algorithms not using Newton-Raphson method}
\label{sec:other methods}

The limiting accuracy derived in Subsection \ref{sec:theoretical limit elliptical} applies to any variant of Newton-Rapshon method for solving KE within a given machine precision, and to any method whose errors are proportional to $f(E)/f'(E)$, since this was the only ingredient used in the demonstration. As we shall discuss in this subsection, many alternative methods that are not based on Newton-Raphson scheme also involve similar errors. Therefore, they are constrained by similar limits on the accuracy. This is not the case of the bisection method, that can attain an accuracy $\mathcal{E}_\text{bisection}\simeq \varepsilon E$.

A first example is Inverse Quadratic Interpolation \cite{Brent1973}. In this case, an iterative solution of KE is computed at the stage $n+1$ as a sum of three terms involving three previous approximations, as
\begin{equation}
    E_{n+1} = \frac{f(E_{n-1})f(E_{n})} {\left[f(E_{n-2})-f(E_{n-1})\right]\left[f(E_{n-2})-f(E_{n})\right]} E_{n-2}+ \cdots 
\end{equation}
where the dots indicate two similar terms that can be obtained from that shown explicitly by cyclical permutations of $E_n$, $E_{n-1}$ and $E_{n-2}$. When the approximation converges to the solution, at least one of the differences $E_n-E_{n'}$ becomes very small so that the corresponding difference of the $f(E_n)-f(E_{n'})$ in the denominator can be approximated by $f'(E_n) (E_n-E_{n'})$. As a consequence, the error will also be proportional, at least, to $f(E)/f'(E)$, thus resulting in a  similar limit to that obtained in Subsection \ref{sec:theoretical limit elliptical}. A similar conclusion can be reached for the secant method \cite{Brent1973}, since it also implies division by differences $f(E_n)-f(E_{n'})$ in close points.

 Other methods, such as inverse series \cite{Stumpff1968,Colwell1993,Tommasini2021}, or splines \cite{Tommasini2020a,Tommasini2020b}, will also be affected by similar limits since they use divisions by $f'$ to compute the coefficients of their expansions. Moreover, CORDIC-like methods for solving the KE  have been proposed and shown to be affected by a limiting accuracy $\sim 10^{-15}\,\sqrt{\frac{2}{1-e}}\,\text{rad}$ in double precision \cite{Zechmeister2018,Zechmeister2021}. This $e$ dependence is similar to that affecting the CNR and MNR methods that we have demonstrated here. 

As we have seen in Subsections \ref{sec:example}, \ref{sec:theoretical limit elliptical}, and \ref{sec:M-dependence}, the bisection method can reach the machine limit $\varepsilon E$ for the accuracy. Since it usually requires a large number of iterations, involving a sine evaluation in each for KE, it is usually much slower than other methods such as the CNR, MNR, or the spline algorithms. A possible way out is to use a faster algorithm for the values of $e$ and $M$ in which it can provide the best accuracy, and switch to the bisection method in the critical region. This idea will be used in Ref. \cite{Tommasini2021b} for the design of 
two routines 
that efficiently solve KE for all values of $e\in[0,1-\varepsilon]$ and $M\in[0,2\pi]$ 
with the best allowed accuracy 
in double precision.  Obviously, this idea can also be applied to the hyperbolic KE.

\section{Conclusions}

The work of RPP \cite{Raposo2017} is a valuable contribution for the excellent seeds they provide for the numerical solution of Kepler's Equation. This improvement enables the MNR algorithm to reach its accuracy limits with a small number of iterations. However, such limits were not recognized properly in Ref. \cite{Raposo2017}, and they were explored here. We hope that our results can also help for a more correct use of the RPP methods. 

To summarize, in this article we demonstrated the following results.

\begin{enumerate}
    \item We disproved the claim of Ref. \cite{Raposo2017} that the accuracy of the CNR and MNR methods with their precise seeds could always be set at the level $10^{-15}$ rad (in double precision) or $10^{-34}$ rad (in quadruple precision) for solving KE for any values of the eccentricity $e$ and the mean anomaly $M$, including in the singular region, corresponding to values of e and M close to 1 and 0, respectively. We argued that such claim contradicts their Equation (20) and their accuracy on $f(E)$, which is declared to be the machine epsilon $\varepsilon$.

    \item     We also proved that the results described in Ref. \cite{Raposo2017} underestimate the correct values of the errors that can actually be obtained with a careful numerical computation using their method, when a scan of the critical region is included.

    \item  We theoretically demonstrated explicit analytical limits,
    given in Equations (\ref{eq:limit-abs-error}) and (\ref{eq:limit-error-e}),
    for the best accuracy that can be achieved, within a given machine precision, with any implementations of the classical or modified Newton-Raphson methods for KE, including those described by RPP. We proved that such expressions describe very well the actual numerical errors. Especially in the case of high eccentricity, these results differ dramatically from the incorrect accuracy given by RPP. The difference can be appreciated in the singular region $e\gtrsim 0.976$ and $M\lesssim 0.014$ rad. In particular, the limiting accuracy diverges as $\sim\varepsilon/\sqrt{2(1-e)}$ for $e\to1$, well above the constant value ($10^{-15}$ rad in double precision) that RPP declare.

    \item The values of the expression $|E-e\sin E -M|$
    that correspond to the limiting accuracy are not of the order $\sim\varepsilon$, as RPP assumed, but $\sim\varepsilon E$. This also implies that the prescription for stopping the iterations given in Ref. \cite{Raposo2017} should be modified in order to obtain the best results their code can provide.

    \item We provided a more general limit that is valid also for the accuracy of schemes that do not use derivatives, like bisection and unlike Newton-Raphson method and its generalizations. Such a universal limiting accuracy is simply given by the expression $\mathcal{E}_\text{abs}\simeq \varepsilon E$. 
    
    \item We demonstrated similar limits for the accuracy of the solution of the hyperbolic KE. If such an equation is solved using the CNR or MNR methods, the limiting accuracy is given by Equations (\ref{eq:limit-abs-error-H}) and (\ref{eq:limit-error-e-H}). In particular, for $e\to1$ the accuracy diverges as $\sim\varepsilon/\sqrt{2(e-1)}$.
    Using non-derivative methods like bisection, the accuracy can be set at the level $\sim \varepsilon H$, where $H$ is the hyperbolic eccentric anomaly.

\end{enumerate}

The methods described in this article can provide guidelines for the design of accurate solutions of the elliptic and hyperbolic Kepler equations, and more in general for studying the accuracy of numerical algorithms for astrophysics. In particular, they will be used 
for the design of 
two routines 
that efficiently solve KE for all values of $e\in[0,1-\varepsilon]$ and $M\in[0,2\pi]$ 
with the best allowed accuracy 
in double precision
\cite{Tommasini2021b}.

\section*{Acknowledgements}

This work was supported by grants 67I2-INTERREG, from Axencia Galega de Innovaci\'on, Xunta de Galicia, and FIS2017-83762-P from Ministerio de Economia, Industria y Competitividad, Spain.

\section*{Data Availability}

No new data were generated or analysed in support of this research.



\bibliographystyle{plain}
\bibliography{comment-mnras.bib} 



\end{document}